\documentclass[%
reprint, 
amsmath,amssymb,
 aps,
]{revtex4-1}

\usepackage{graphicx}
\usepackage{dcolumn}
\usepackage{bm}
\usepackage{amsmath}
\usepackage{float}
\usepackage{subfigure}
\usepackage{comment}
\newcommand{\norm}[1]{\left\lVert#1\right\rVert}

\begin{document}




\title[Submitted Manuscript]{Role of Rabi oscillations in radiative states due to the fully absorbing smaller plasmonic nanoparticles}


\author{Kritika Jain}
\author{Murugesan Venkatapathi}%
\email{murugesh@iisc.ac.in}
\affiliation{Computational and Statistical Physics Laboratory, Indian Institute of Science, Bangalore, 560012}




\begin{abstract}
The modified radiative and non-radiative states due to the weak coupling of an emitter with other resonant objects (Purcell effect), can be recast as a quantum interference of the paths of the photon that define the classical scattering and absorption by the object. When the coupling is stronger, additional paths representing the (Rabi oscillations or) possible re-absorption of the photon from the excited object, by the emitter at ground-state, have to be included in the quantum interference. The effect of these additional Rabi paths of the photon on the radiative states and the efficiency of spontaneous emission, can be approximated using a simple one-loop correction to the weak-coupling approximation. This effect is especially evident in the anomalous enhancements of emission due to extremely small non-scattering (or fully absorbing) metal nanoparticles less than 10 nm in dimensions. Extending these corrections to a collective model of spontaneous emission that includes multiple emitters and such very small metal nanoparticles coupled to each other, the large contribution of Rabi paths to radiative decay in such bulk materials is elucidated.
\end{abstract}

\maketitle

\section{Introduction}
The role of vacuum modes on spontaneous emission of photons has been well elucidated and the reversible exchange of a photon between a mode of vacuum and an atom inside a micro-cavity is observable \cite{zhu1990vacuum, bernardot1992vacuum, boca2004observation, yoshie2004vacuum}. This Rabi oscillation of the excitation between the cavity and an emitter is the sign of a strongly coupled emitter-vacuum system, and the coupling strength inferred by the ratio of the frequency of oscillation and the decay rate of the cavity, is large \cite{Chikkaraddy2016strong}. In the absence of the cavity i.e. weak coupling of the emitter with vacuum, when a proximal resonant object is introduced, such Rabi oscillations with the object can emerge indicating a strong coupling with it \cite{bellessa2004matter, zengin2015matter, Schlather2013splitting}. The modified decay rates in such cases is determined by the additional optical modes available for the spontaneous emission \cite{Vats2002LDOS}. The strong coupling of emitters with metallic nanostructures are of interest in quantum sensing and information processing \cite{otten2015entanglement, mathieu2020absorption, Xavier2021quantumsensing}, and the strength of coupling possible with such metallic dissipating systems has been studied in detail \cite{Heiko2018coupling, pelton2019strong, manish2020absorption}.

Even in this case of a strong coupling, the partition of optical states into the radiative and non-radiative parts reflected the classical absorption and scattering properties of this object \cite{chang2007strong, van2012spontaneous, delga2014quantum}, as is the case in the weak-coupling (Purcell) regime. Recently, it was shown that this conventional partition can result in significant anomalies when the emitter is coupled to fully absorbing matter i.e. non-scattering very small metal nanoparticles less than 10 nm in dimensions, and also in the enhancements expected in surface-enhanced-Raman-spectroscopy (SERS) \cite{jain2019strong, dutta2019large}.  Observations of such enhancements of spontaneous emission near these fully absorbing metal nanoparticles have been reported for more than a decade \cite{haridas2010controlled,haridas2010photoluminescence,kang2011fluorescence,haridas2013photoluminescence}. It should be noted that in the very small metal particles with dimensions approaching the mean free-path of the electrons, a broadening and minor shifts of the resonance along with oscillatory spectra are expected due to the scattering of electrons at the boundaries of the particle \cite{Fujimoto1968Early_EELS, Kreibig1995, Scholl2012NP_EELS}. But the modified effective permittivity due to these dissipative processes in the very small metal nanoparticles does not explain the enhanced radiative decay from the coupled emitters \cite{agarwal1983NonlocalSphere, Ginzburg1984, Leung1990NonlocalResonance}. In the case of SERS, a larger absorbing resonant metal nanostructure increases the radiation exciting a molecule in its near-field by orders of magnitude, but surprisingly without apparent absorption of the emitted photons. This divergence of SERS from first principle theoretical models has increased during which the reported SERS enhancements have grown from $10^4$ to $10^{14}$ \cite{schatz2006electromagnetic, Moskovits2013review, kneipp2016SERSrigorous, Heeg2020SERS}. Thus, a revisit of the radiative and non-radiative optical states due to a nanostructure may be required for the case of moderate to strong coupling-strengths.

We begin with a phenomenological description of the strong-coupling regime of the emitter and a nanostructure, to deduce a simple correction to the conventional partition of optical states into radiative and non-radiative parts. This renormalization of the radiative and non-radiative parts invokes the emitter's possible re-absorption of the photon from the excited dissipating structure, to include an one-loop correction \cite{jain2019strong}. It can be equally described as the result of non-Markovian interaction between the emitter and the proximal absorbing object, where the decay is not a simple superposition of two independent decay processes from the emitter and the excited object. The effects of non-Markovian interactions on the \textit{total} decay, and its dynamics, have been studied before for both cavities and a metal surface \cite{Gonzalez2010Markov, Lodahl2011Markov, agarwal2013quantum}. A recent work on the effect of non-Markovian interaction between an emitter and a metal nanoparticle, on the radiative and non-radiative parts of the decay, validates this one-loop correction \cite{jain2021nonMarkov}. The one-loop correction for the efficiency of emission is especially useful for studying bulk materials with many emitters and the smaller metal nanostructures interacting with each other, and is the main goal of this work. In this paper, we extend this correction of the radiative and non-radiative decays to a collective model of many coupled emitters and metal particles, and use it to study this effect in such materials.

One can recast the modified spontaneous emission due to a body as a quantum interference of all the possible paths of the emitted photon \cite{macovei2007strong,shatokhin2005coherent,safari2019plasmon-phase}. The resulting effects on the radiative and non-radiative decay channels can be explained succinctly using Figure 1. When we include the Rabi paths of the photon, which include the re-absorption of a photon by the emitter that would have otherwise dissipated in the object, it permits a fully absorbing non-scattering object to increase the efficiency of spontaneous emission. Let $\Gamma^r_o$ and $\Gamma^{nr}_o$ be the known radiative and non-radiative decay rates of the isolated emitter in vacuum, adding to $\Gamma_o$. Similarly, $\Gamma^r$ and $\Gamma^{nr}$ add to the total metallic contribution $\Gamma$. The total radiative and non-radiative rates of the system are a sum of the free-space and metallic components.

At large relative separations, the emitter is uncoupled and we observe no effect of the object on the emission. When the resonant object is relatively close to the emitter, the increase in coupling strength with Rabi frequency $\Omega \sim \Gamma_0$, allows the interference of classical paths A' through the object and the direct free-space path A shown in Figure 1a. But note that this path A' is dominant only when the decay from the object is significantly faster than the Rabi oscillations ($\Gamma \gg \Omega$). The above condition signifies the weak coupling of the emitter with the object. It is known that the interference of these two paths A' and A at all points P, given by the superposition of the scattered field from the object and the direct field from the emitter, represents the self-interaction, additional radiative and non-radiative decay due to the object.

\begin{figure}
		\centering
		\includegraphics [width=1.0\linewidth]{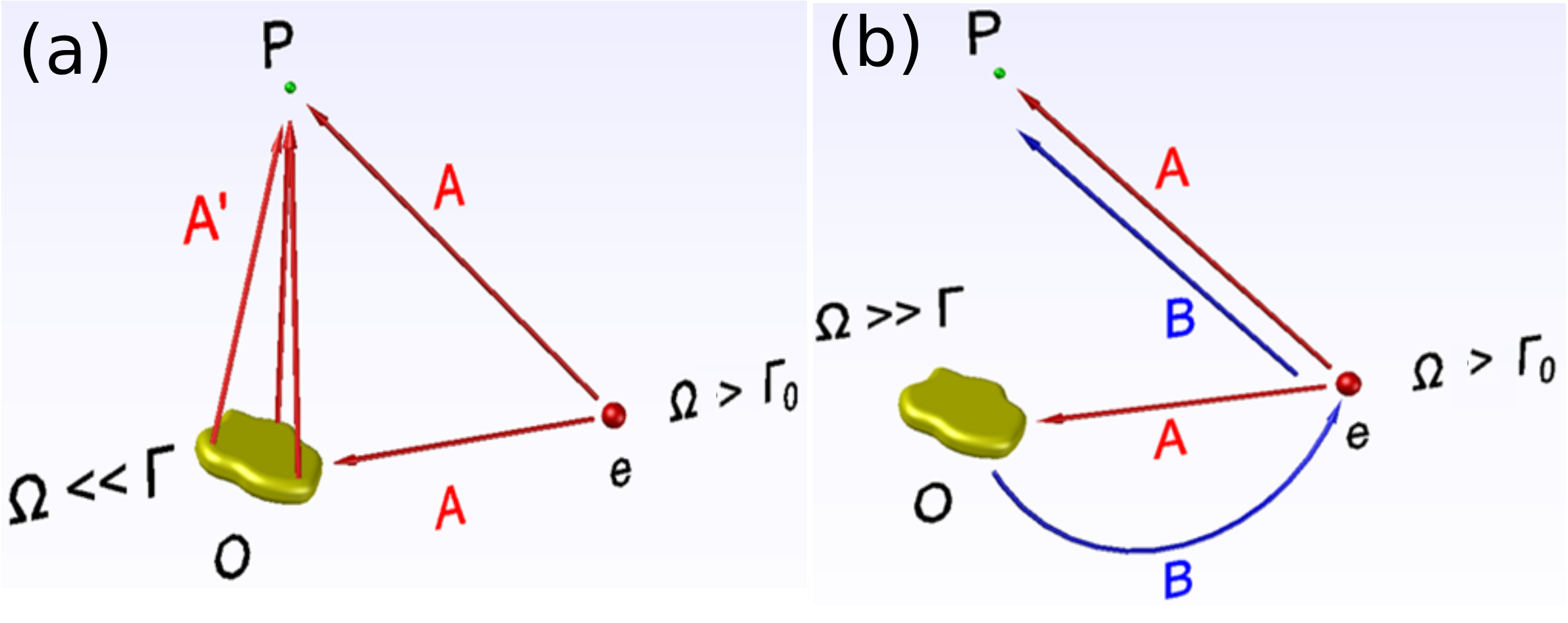}

		\caption{Coupling of emitter $e$ with object $O$ and resulting coherent paths of the photon to a point P; $\Omega$ is the frequency of emitter-object Rabi oscillations, $\Gamma$ and $\Gamma_o$ are the decay rates from the object and the emitter in free-space respectively. (a) Interference of paths A and A' results in possible additional decay due to the object. In classical electromagnetics, it represents absorption and scattering due to the interference of incident field of the emitter and scattered field of the object. (b) When the coupling with the object is stronger the photon returns to the emitter through paths B (in blue), as the excitation is limited to a quantum of energy and the self-interaction or total decay can be oscillatory. Also, the paths A' vanish along with the distinction between scattering and absorptive properties of the particle. Effect of quantum interference of paths B and A on the radiative and non-radiative parts of the decay can be approximated using an one-loop correction.}
		\label{Figure1}
\end{figure}

However, when the emitter is strongly coupled to the object and we consider only a \textit{single} Rabi oscillation, paths A' become irrelevant as Rabi oscillations are much faster than the decay in the object ($\Omega \gg \Gamma$). The alternate path of the photon through the object is now path B, and its interference with A as shown in Figure 1b should be of primary interest. The crucial difference in this path is that there is no decay in the object and hence no absorption. When $\Gamma \sim \Omega$, the corresponding probabilities of these mutually exclusive paths (1a and 1b) are given by $1-e^{-\frac{\Gamma}{\Omega}}$ and $e^{-\frac{\Gamma}{\Omega}}$; as recalled in equation \eqref{eq:probability} later.

In these cases, the decay rates $\Gamma^r$ and $\Gamma^{nr}$ of a metallic object evaluated using only the paths A' can be corrected to evaluate the actual quantum efficiencies of emission, as a perturbation. Note that $\Gamma$, $\Omega$ $\ll \omega_0$, the resonance frequency of emitter, in the regime of interest here. The decay through path B where dissipation is absent, replaces non-radiative decay of the object's dipole mode numbered `1' with radiative decay \cite{jain2019strong}. Using the probability of this path B given by $e^{-\frac{\Gamma}{\Omega}}$, the renormalized partition of decay rates was given by:
\begin{equation} \label{Gamma_leak}
\Gamma_{leak}=e^{-\frac{\Gamma}{\Omega}}\cdot\Gamma^{nr}_1
\end{equation}

The effective decay rates are:
\begin{align}  \label{decay_effective}
\Gamma^r_{eff} &= \Gamma^r_o+\Gamma^r+\Gamma_{leak}\\
\Gamma^{nr}_{eff} &= \Gamma^{nr}_o+\Gamma^{nr}-\Gamma_{leak}
\end{align}
In the above, the correction to the decay rates becomes significant with $\Gamma_{leak} \gtrsim \Gamma^r$, only when $\Omega \gtrsim \Gamma$ or $\Gamma_1^{nr} \gg \Gamma^r$ in \eqref{Gamma_leak}. The latter condition is relevant to an emitter very proximal to a metal nanostructure as in SERS, where very large non-radiative decays and extremely low quantum efficiencies are predicted by the conventional partition of optical states. They underestimate the possible gains severely, even when the repeated excitation of the molecule in the intense near-field near the metal and the increased ground-state probabilities (i.e. large total decay rates) of emitter are taken into account. Whereas this corrected efficiency of emission well-predicts the large gains observed in experiments \cite{jain2019strong}. In the case of fully absorbing metal nanoparticles, $\Omega \gtrsim \Gamma$ representing a moderate coupling strength is also satisfied, even when reasonably well separated from the emitter. The Rabi paths and the oscillatory self-interaction of the emitter does make the dynamics of emission non-Markovian i.e. exponentially damped oscillatory decay for a single excitation, which can manifest as a multi-exponential decay in ensembles \cite{Lodahl2011Markov, agarwal2013quantum}. Since the above one-loop correction is directed at the weak to moderate coupling strengths as in our examples, and we are interested in the time-averaged efficiencies of emission, we ignore these marginal effects.

\section{Methods}
The significant point of interest here is the above effect of Rabi paths when multiple emitters are strongly coupled to metal nanoparticles. This is practically significant as bulk materials have many potential emitters, and even for a single emitter like quantum dot the effect of its finite size on the coupling has to be considered \cite{dutta2019large}. Of specific interest are the extremely small metal nanoparticles which have a negligible scattering efficiency and are fully absorbing; see the appendix for the variation of the strength of coupling with size of metal nanoparticles. In the rest of this paper we use a model of many coupled dipole emitters and metal particles, both with and without including the Rabi paths described above, to elucidate the density of optical states in such materials.

\begin{figure}[h!]
		\centering
		\hspace*{-2mm}
		\includegraphics [width=1.0\linewidth]{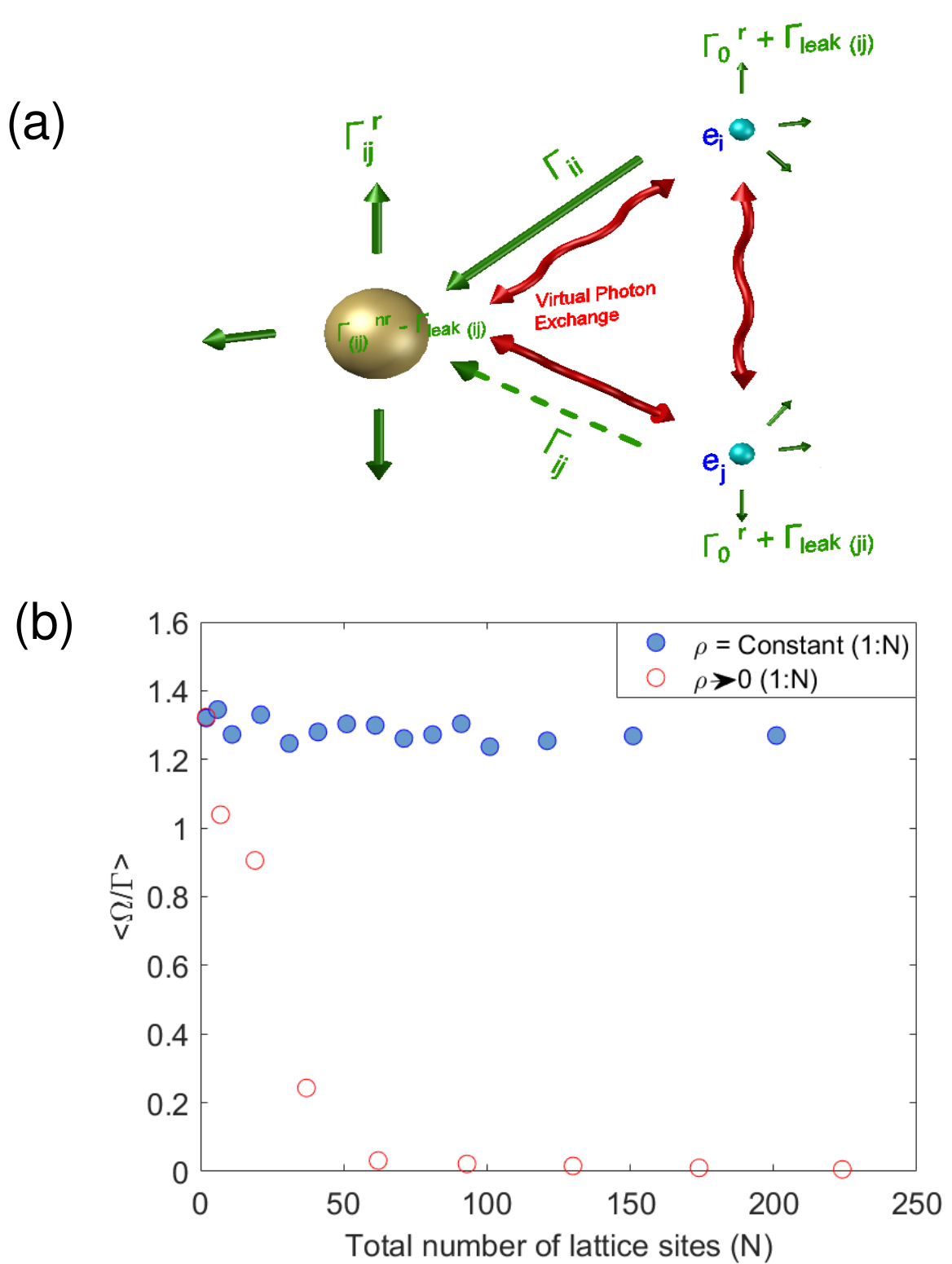}
		\caption{(a) The one-loop corrections extended to multiple emitters $i$ and $j$; it includes Rabi oscillations among emitters, and also between an emitter and the metal nanoparticle. (b) The expected coupling strengths with the metal nanoparticle in the Markovian approximation, evaluated by averaging $\langle J|K|J \rangle$ over modes $J$ in many configurations for a given $N$.}
		\label{Figure3}
\end{figure}

A coupled system of Lorentz dipole oscillators was used to model an excitation of one quantum of energy shared among identical emitters. It represents weak excitations given by one excited emitter among the $N$ possible emitters i.e. superpositions of one excited emitter and $N-1$ other emitters at ground state \cite{svidzinsky2010cooperative, Agarwal2011Wstates, jin2011photon}. The coupling of this ensemble of emitters with a metal particle (Figure \ref{Figure3}a) as required for the study here, is included by the modification of the Green tensors coupling the emitters due to the addition of the nanoparticle. This model was first treated analytically under long-wavelength approximations for a spherical metal particle \cite{pustovit2010plasmon}, was extended to retarded waves using computational methods \cite{van2012spontaneous}, and further to arbitrary geometries using discrete dipole approximations such as in this work \cite{venkatapathi2014collective}. The latter numerical methods allow us to accurately compute the retarded Green tensors coupling any given pair of emitters in an arbitrary system containing emitters and other nanostructures; see \eqref{eq:total-decay} for example. The pair-wise self-energy contribution of $N$ coupled Lorentz dipole oscillators proximal to metal nanostructure is:
\begin{align}\label{self_energy_multipole}
\Sigma^{total}_{jk}(\omega) &= \frac{-2\pi e^2\omega}{mc^2}\mathbf{e}_j \cdot \mathbf{G} (\mathbf{r_j,r_k};\omega) \cdot \mathbf{e}_k - \delta_{jk}\frac{i\Gamma_o}{2}\\
&=\Delta_{jk}^{total} - \frac{i\Gamma_{jk}^{total}}{2}
\end{align}
and the above can be integrated over polarization vectors $\mathbf{e}$, and with a line-shape around the emission frequency $\omega_0$ if required. Here $e$ is the oscillating charge, $m$ is its mass, and $c$ is speed of light. The decay rate is given by the imaginary part where the reduced Planck's constant was divided out of the self-energy in \eqref{self_energy_multipole}. $\Delta_{jk}^{total}$ and $\Gamma_{jk}^{total}$ are real and imaginary parts of the self-energy and represent entries of $N \times N$ matrices. Green dyadic $\mathbf{G}$ represents interaction between the two dipole emitters in the presence of the metal nanostructure. The diagonal entries of the above matrix given by $j=k$ represent the total self-interaction of a single emitter due to the nanoparticle and vacuum. This model represents a dipole approximation of two-level emitters in the weak vacuum-coupling regime, and the Fermi golden rule can be used to relate the decay rates to the density of optical states \cite{Tignon95Fermi, Debierre2015Fermi}. 

This widely used conventional description of the self-interaction in \eqref{self_energy_multipole} includes an implicit rotating wave approximation ($|\Delta^{total}_{jk}| \ll \omega$). Note that this real part of self-energy provides us the energy split between the two modes of the strongly coupled oscillators $j$ and $k$, and corresponding Rabi frequency, in this approximation \cite{pustovit2010plasmon,schulzgen1999direct}.

\begin{equation}
   \delta E_{jk}/\hbar = \Omega_{jk} = 2|\Delta^{total}_{jk}| 
\end{equation}
where $\hbar$ is the reduced Planck's constant. The Rabi frequency $\Omega$ can be further generalized to include the effects of dephasing due to asymmetry in the damping of the two oscillators  \cite{agarwal1985vacuum,kamada2001exciton,pelton2019strong,alpeggiani2012surface}. The self-energy matrix evaluated in \eqref{self_energy_multipole} can be further decomposed into its metallic contribution $\Sigma$, and $\Sigma^0$ the contribution due to direct interaction among the $N$ coupled dipole emitters.
\begin{equation}
\Sigma^{total}_{jk}(\omega) = \Sigma^0_{jk}(\omega) + \Sigma_{jk}(\omega)
\end{equation}
These are given by a corresponding decomposition of the Green tensors $\mathbf{G}$ coupling two emitters $j$ and $k$ in the presence of the metal nanoparticle.
\begin{equation} \label{eq:total-decay}
 \mathbf{G}(\mathbf{r_j,r_k};\omega) = \mathbf{G}_0(\mathbf{r_j,r_k};\omega) - \hat{G}_{jp}\hat{G}_{pq}^{-1}\hat{G}_{qk}
\end{equation}
where the 1.9 nm radius metal particle was in turn decomposed finely into many (552) dipole grains referred by indices $p$, $q$ arranged in a hexagonal close packed form, each representing a spherical sub-volume 0.19 nm in radius. This numerical approach allows us to include retarded interactions in a system of dimensions comparable to the wavelength of emission. The largest size of the grains allowed is determined by the wavelength of emission and material properties of the nanoparticle, but they can be smaller, and the number of such polarizable oscillators $m$ can be increased for a higher accuracy to include the effect of its higher order modes. The global matrices containing the Green dyads as $3\times3$ blocks are given by
\begin{align}
\hat{G}_{pq}(3p\rightarrow3p+2,3q\rightarrow3q+2)&=\mathbf{G}_0(\mathbf{r_p,r_q};\omega) \text{ for } p \neq q\\
\hat{G}_{jp}(0\rightarrow2,3p\rightarrow3p+2)&=\mathbf{G}_0(\mathbf{r_j,r_p};\omega)\\
\hat{G}_{qk}(3q\rightarrow3q+2,0\rightarrow2)&=\mathbf{G}_0(\mathbf{r_q,r_k};\omega)
\end{align}
where the indices $p$, $q$ = $0, 1,2,\dots m-1$ represent oscillators in the particle. Thus, $\hat{G}_{jp}$ and $\hat{G}_{qk}^T$ are of size $3\times3m$, and $\hat{G}_{pq}$ is of size $3m\times3m$. The vacuum self-interaction of grains in the metal particle are given by their polarizability tensors $\hat{\alpha}$ for a general case of non-isotropic polarizability. Here, we use the scalar isotropic polarizability of the metal, $\alpha$, determined from its size and the dispersive permittivity of the bulk material, using the Clausius-Mosotti relation \cite{purcell1973DDA} and its corrections for any dispersion by the lattice of grains \cite{draine1994discrete}. When $p=q$, the diagonal entries of these $3\times3$ dyads are given by $1/\alpha$. The self-interaction dyads of emitters in the background $\mathbf{G}_0(\mathbf{r_k,r_k};\omega)$ are set to $\mathbf{0}$, as $\delta_{jk}\frac{i\Gamma_o}{2}$ has been explicitly included in the self-energy matrix of \eqref{self_energy_multipole}. The other dyads $\mathbf{G}_0$ for the interactions are calculated using the solutions of point source in a homogeneous background:
\begin{equation}
\bigtriangledown \times \bigtriangledown \times \mathbf{G}_0(\mathbf{r,r_j};\omega)- k^2\mathbf{G}_0(\mathbf{r,r_j};\omega) = \mathbf{I}\delta(\mathbf{r-r_j})
\end{equation}
where $\mathbf{I}$ is a unit dyad, the wave number $k=\sqrt{\epsilon} \frac{\omega}{c}$, and $\delta(\mathbf{r-r_j})$ represents the point source. This gives us the required dyadics for the interaction among the point-dipoles:
\begin{equation} \label{dipole_green}
\mathbf{G}_0(\mathbf{r_i,r_j};\omega)=(\mathbf{I}+\frac{\bigtriangledown\bigtriangledown}{k^2})g(\norm{\mathbf{r_i}-\mathbf{r_j}})
\end{equation}
where \(g(r)=\frac{e^{ikr}}{4\pi r}\). With the self-energy matrix in \eqref{self_energy_multipole}, the eigenstates $|J \rangle$ of the coupled many emitter-metal system are calculated using:
\begin{equation}
\Sigma^{total} |J \rangle =\Delta_J - i \frac{\Gamma_J}{2} |J \rangle
\end{equation}

Normalized eigenvectors $|J\rangle$ represent one of $N$ collective modes of emission here. The imaginary part of an eigenvalue, $\Gamma_J$, represents the total decay rate of the mode while the real part $\Delta_J$ represents the energy shift. The energy shifts $\delta E_J$ and Rabi frequency $\Omega_J$ in the collective mode are:
\begin{equation}
\Omega_J=2\delta E_J/\hbar=2|\Delta_J|
\end{equation}

Note that $|J \rangle$ is not an eigenstate of $\Sigma$ that represents only the metallic contribution. We evaluate contributions of the metal to the energy shifts and decay rates of a mode using an entry-wise decomposition of $\Sigma_{jk}$ into $\Delta_{jk} - i\Gamma_{jk}/2$ as the real and imaginary parts, and corresponding expectations $\langle J|\Delta|J \rangle$ and $\langle J|\Gamma|J \rangle$. Each component of matrix $\Delta_{jk}$ represents the rate of Rabi oscillations between two dipoles at position $\mathbf{r_j}$ and $\mathbf{r_k}$ through the metal nanostructure, and matrix $\Gamma_{jk}$ represents effects of metal couplings on decay rates of the two dipole emitters. The strength of coupling between emitters through the metal is given by $K_{jk}=2|\Delta_{jk}|/\Gamma_{jk}$ and its expectation $\langle J|K|J \rangle$ averaged over $|J \rangle$ represents the significance of metallic Rabi paths of the photon for the system (such as the example in Figure \ref{Figure3}b).

The decay components $\Gamma_{jk}$ due to the metal can be decomposed into its radiative $\Gamma^r_{jk}$ and non-radiative $\Gamma^{nr}_{jk}$ by a factorization of $\hat{G}_{pq}$ in \eqref{eq:total-decay} using real and imaginary parts of polarizability of dipole granules, which represent the metal nanoparticle \cite{venkatapathi2014collective}. This allows us to use expectations $\langle J|\Gamma^r|J \rangle$ and $\langle J|\Gamma^{nr}|J \rangle$ for the radiative and non-radiative components of the mode. Alternatively, the conventional integral of the outgoing momentum of the collective field of all polarizations given by $|J \rangle$ over an enclosing surface, provides the radiative decay for the mode. The evaluated decay rates thus take into account the effect of the additional paths among emitters and to vacuum, due to the possible absorption and subsequent radiative decay from the metal particle. Note that a larger highly scattering plasmonic nanoparticle increases the radiative decay of a collection of emitters, just as in the case of a single emitter.

Now, the correction to the radiative and non-radiative decay rate components when considering the survival of the photon without a decay for one Rabi oscillation through the metal particle, is given by entries of a matrix $\Gamma^{leak}_{jk}$. These corrections are evaluated from the above described Markovian decay rates as a perturbation. Beginning with the integral of the memory-less probability $density$ function $\Gamma e^{-\Gamma \tau}$ in time $\tau$ for the interval [0,t], its complement gives us the probability of the excited state as $e^{-\Gamma t}$ for all $t$. Restricting the decay in the metal to the interval of one Rabi oscillation as shown below, its complement give us the probability of one exchange of the photon between emitters $j$ and $k$ through the metal particle without a decay.

\begin{equation} \label{eq:probability}
     \mathcal{P}_{exch.} = 1 - \int_{\tau=0}^{\tau=1/2|\Delta_{jk}|} \Gamma_{jk} e^{-\Gamma_{jk} \tau} d\tau
\end{equation}
Note that the decay in the smaller metal particles is slower, and the Rabi oscillations can survive. This exchange of the photon cannot be distinguished from the exchange between the emitters through a radiative decay from the excited nanoparticle, and the resulting correction of radiative and non-radiative matrices is given by \cite{jain2019strong}: 
\begin{equation}
\Gamma^{leak}_{jk}=\mathcal{P}_{exch.}\Gamma^{nr(1)}_{jk}=e^{-\frac{\Gamma_{jk}}{2|\Delta_{jk}|}}\cdot\Gamma^{nr(1)}_{jk}
\end{equation}

Here the superscript (1) refers to the contribution of the dipole mode of the metal nanoparticle representing its coupling to vacuum. The diagonal entries of this matrix reduce to the effect of Rabi oscillations between a single emitter and the metal particle, as given by equation \ref{Gamma_leak} in the introduction. The mode-wise $\Gamma_{leak}$ is calculated by the expectation $\langle J|\Gamma^{leak}|J \rangle$. The effective radiative and non-radiative decay rates remain as given in equation \eqref{decay_effective}, and the corrected values represent a collective mode here. The quantum efficiency of a mode and the expected quantum efficiency are given by:
\begin{align} \label{quantum_efficiency}
Q_J &=\frac{\Gamma^r_{eff}}{\Gamma^r_{eff}+\Gamma^{nr}_{eff}} & Q &= \frac{1}{N} \sum_{J=1}^{N} Q_J
\end{align}

\section{Results}
For our analysis of results, let $\rho$ be the average measure of the coupling of emitters with metal nanoparticles, and it can be defined as:
\begin{align} \label{rho_defn}
\rho = \frac{n_{nc}+n_{c}}{l_{nc}+l_{c}+n_{nc}+n_{c}}
\end{align}

where $n$ represents the number of possible paths of the photon through the metal particle, and $l$ represents paths involving only the other emitters, recalling the relationship of the additional paths to the additional optical states. The subscripts $c$ and $nc$ indicate the classical and the (non-classical) Rabi paths respectively. The decomposition  $\rho = \rho_{c}+\rho_{nc}$ is useful here where they include either $n_{nc}$ or $n_c$ respectively in the numerator of equation \eqref{rho_defn}. $\rho_{nc}$ is a measure of the paths through metal nanoparticles due to Rabi oscillations, while $\rho_{c}$ is a measure of the emitter's weak coupling and further decay from the metal particle. In the conventional evaluations $\rho_{nc}=0$, while in proposed one-loop correction both $\rho_{c}$ and $\rho_{nc}$ are non-zero. We investigate two conditions (a) $\rho$ = constant and (b) $\rho \longrightarrow 0$ in each of two cases; case I: $\rho_{nc}=0$ and case II: $\rho_{nc} \neq 0$. In both cases I and II, the condition (a) $\rho$ = constant demonstrates the variation of local density of optical states (LDOS) due to the metal particle, with increasing number of point emitters that share the excitation in a confined area. From a practical point of view, it relates to finite size emitters like quantum dots with a distributed polarization density coupled to a metal particle. While the condition (b) $\rho \longrightarrow 0$ represents the increase in the number of emitters with a constant area density. This condition demonstrates the smooth convergence of the radiative and non-radiative LDOS near the metal particle to the density of optical states (DOS) of a regular infinite lattice. It is also shown by introducing multiple metal particles as a fraction of this large lattice, that the DOS in the modified lattice is well represented by the radiative and non-radiative LDOS near a metal nanoparticle given by condition (a). This may be relevant for the enhanced emission efficiency possible in bulk materials. A description of the geometries used in these two conditions is given below.

\begin{figure}[h!]
    \begin{subfigure}{}
		\includegraphics[width=.7\linewidth]{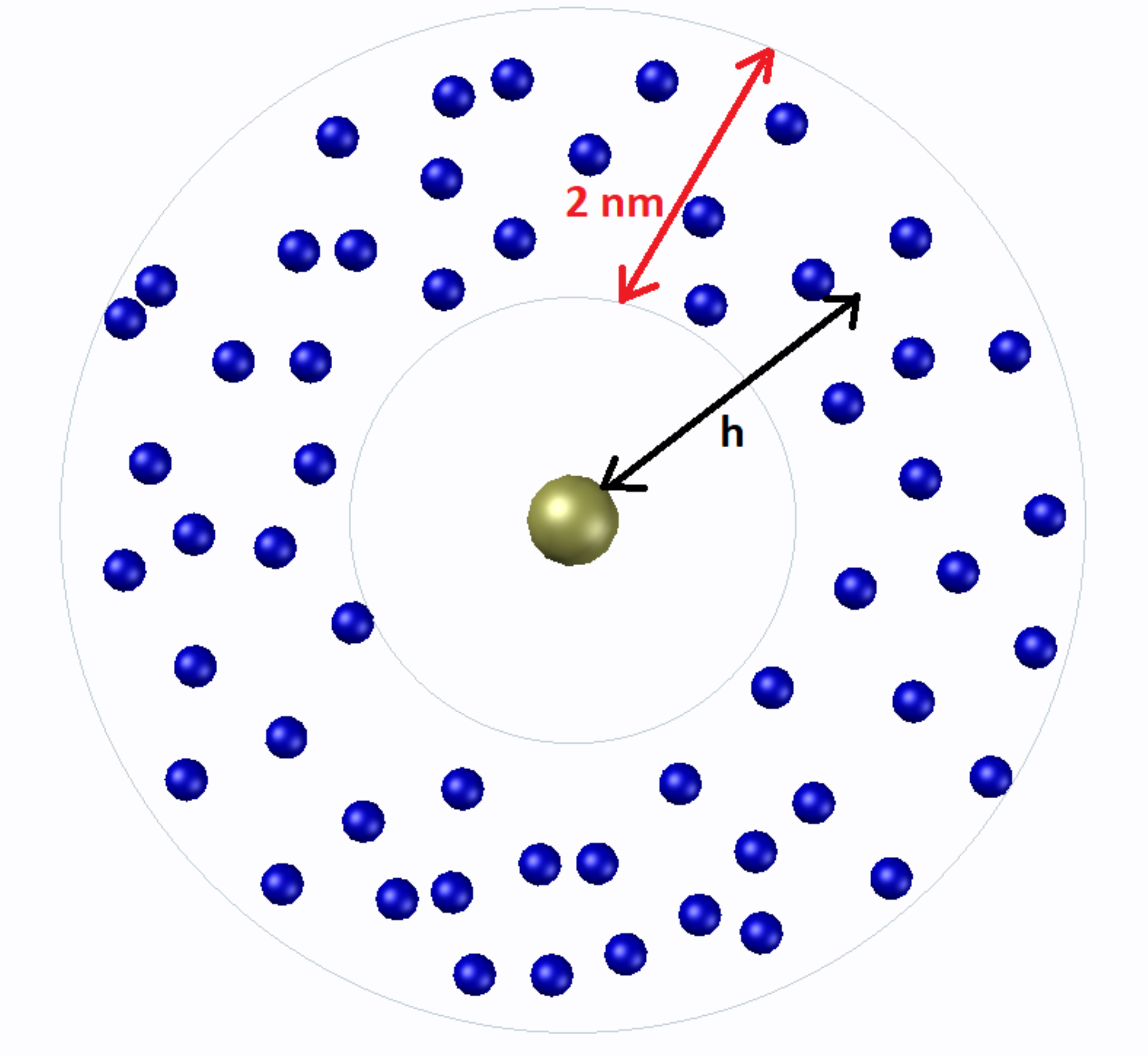}
    \end{subfigure}
    \begin{subfigure}{}
		\hspace*{7mm}
		\includegraphics[width=.7\linewidth]{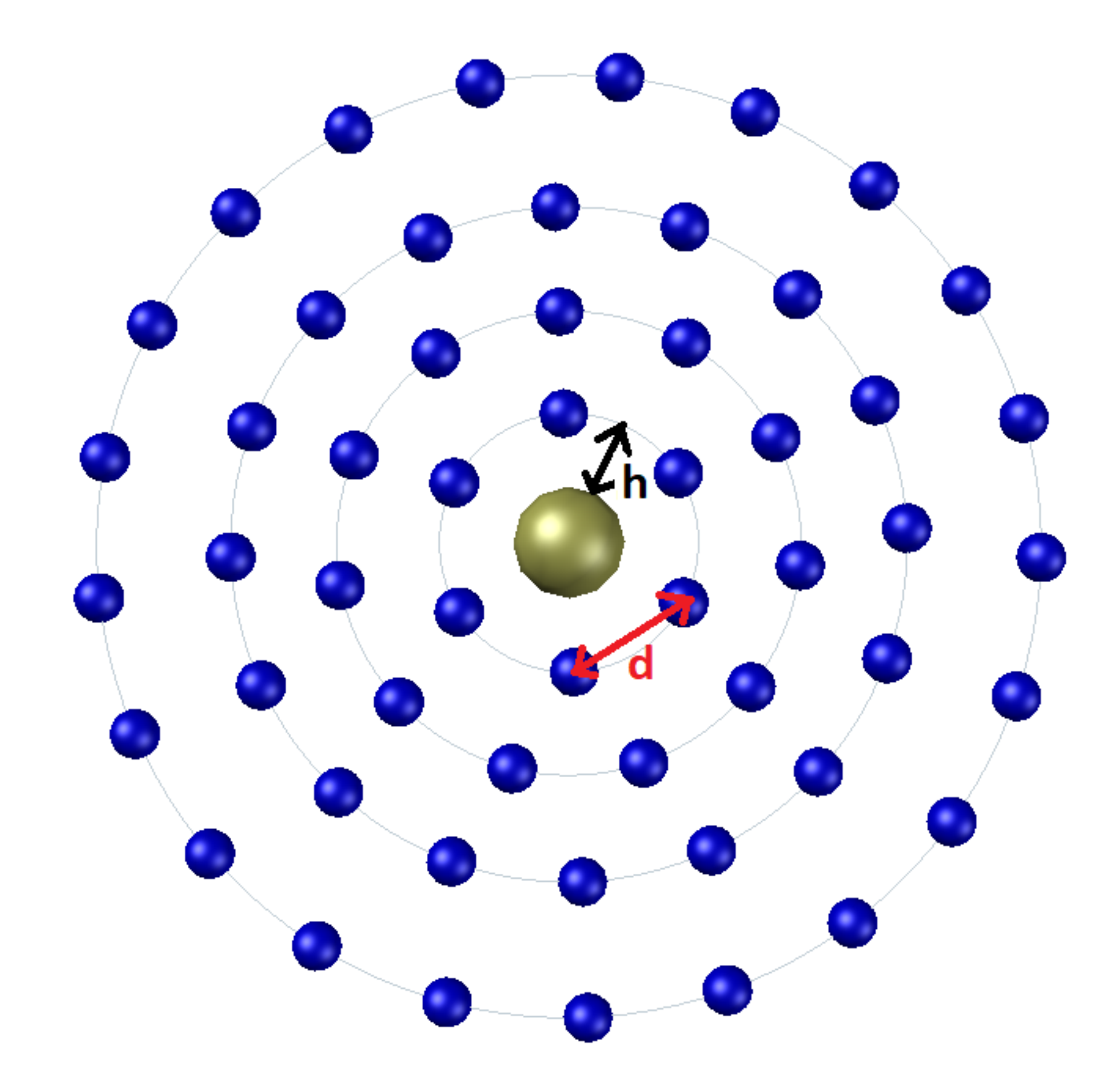}
    \end{subfigure}
	\caption{The larger (golden) sphere is a gold nanoparticle of radius 1.9 nm, which is represented in numerical evaluations using 552 spherical sub-volumes each 0.19 nm in radius. The smaller (blue) spheres represent point emitters where $h$ is distance between nearest emitters and the nanoparticle. Top: (a) Emitters distributed with a constant area: $\rho \approx$ constant with increase of sites. Bottom: (b) Emitters distributed with a constant area density: $\rho \rightarrow 0$ with increase of sites.}
		\label{FigureS2}
\end{figure}

\subsection{Model description}
Here we explain the details of geometries corresponding to $\rho =$ constant and $\rho \to 0$ used for simulations in figures \ref{Figure3} and \ref{Figure4} of the paper. The first geometry represents the case of an excitation localized in a fixed finite area. The second geometry $\rho \to 0$ mimics a lattice of potential emitters extending to infinity, placed around a single metal nanoparticle. The emission wavelength of emitters is 560 nm in free-space and the refractive index of surrounding medium is 1.5. $Q_o = 0.33$ was assumed, and gold nanoparticles of diameter 3.8 nm are coupled to many dipole emitters. Each data point in the figures represents the mean of a number of simulations of many random configuration of polarizations and permuted positions, numbering greater than $N$ until the variance of the computed values is much smaller than its mean.

\subsubsection{Constant area : $\rho = constant$} \label{case:A}

Here we consider $N-1$ dipole emitters around a metal nanoparticle of diameter 3.8 nm at an average distance of h=7.6 nm from the surface of the nanoparticle as shown in Figure \ref{FigureS2}. The point dipole emitters are uniformly distributed in a fixed area of shell of 2 nm. The decay rates are computed with increasing number of emitters in the shell to observe the effect of Rabi paths with metal nanoparticle. The quantum efficiency is roughly constant with the increase of number of emitters as shown by constant quenching in Figure \ref{Figure4} when limited to the classical paths through the metal particle, and constant enhancement in \ref{Figure4} due to inclusion of the Rabi paths through the metal nanoparticle. Note that paths $l$ that include only emitters, and classical paths involving metal $n_c$ increase as a factorial of $N$. But so does the paths $n_{nc}$ due to Rabi oscillations between the metal and emitters, which is relevant for case II. The latter is possible as the stronger coupling of all emitters with the metal is ensured in this geometry. This ensures that the fraction of metallic paths $\rho \approx$ constant.

\subsubsection{Constant area density : $\rho \to 0$} \label{case:B}
Here we consider $N-1$ dipole emitters around a metal nanoparticle of diameter 3.8 nm, where distances among emitters are fixed so that number of emitters per unit area i.e. area density is constant (see Figure \ref{FigureS2}). The first emitter is placed at a distance of h=7.6 nm from the surface of the metal nanoparticle and then additional emitters are placed on lattice sites which are at $d$ $\approx$ 9.5 nm apart from each other. The lattice sites are located in concentric circles around metal nanoparticle and distance between them is chosen so that first circle around metal contains exactly 6 emitters which may represent a hexagonal lattice.  Note that the Rabi paths $l$ involving only emitters increases as a factorial of $N$, while the paths $n_{nc}$ through Rabi oscillation with metal marginally increases with $N$ up to a constant, beyond which emitters are not coupled strongly enough to the metal. The classical paths $n_c$ due to a weak coupling increase as a factorial of $N$ initially, but as the couplings reduce further it converges to a constant when the farther emitters are not sufficiently coupled to the metal particle. This results in the fraction of metallic paths $\rho \to 0$ as $N$ increases.

But both $\rho$ and $\rho_{nc}$ increase when some of the emitters are  randomly replaced by metal nanoparticles on same lattice sites so that overall metal to lattice sites ratio is 1:6. The green squares in Figure \ref{Figure4} show this qualitative behaviour and these simulations include 224 lattice sites out of which 37 are metal nanoparticles. Note that we used dipole metal nanoparticles to model this special case because of computational complexity of numerous multipole nanoparticles, and this underestimates the quenching and marginally overestimates the efficiency of emission.

\subsection{Discussion}
\begin{figure}[h!]
		\centering
		\includegraphics [width=1.0\linewidth]{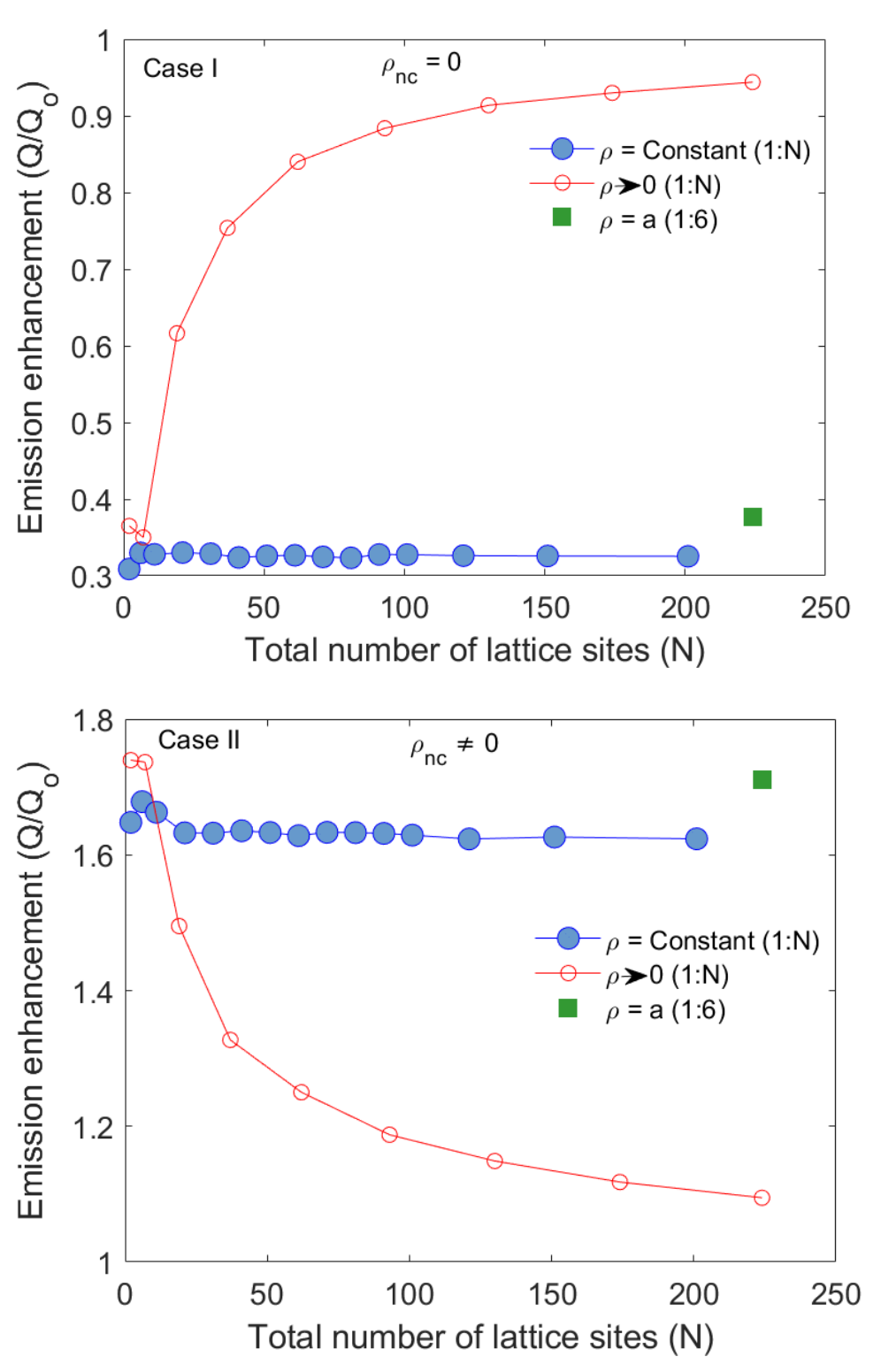}
		\caption{Modified emission due to gold nanoparticles 1.9 nm in radii, free-space emission wavelength $\lambda$ = 560 nm and $Q_o$ = 1/3. The green squares show that replacing some emitters with metal nanoparticles in the system of constant area density i.e. $\rho \to 0$, retrieves the properties of the system with $\rho$=Constant. Case I: quenching when only classical paths of the metal are included. Case II: enhancements when Rabi oscillations with metal are also included.}
		\label{Figure4}
\end{figure}

In case I, the Rabi oscillations between the emitters are accounted in the superpositions but the Rabi oscillations with the metal particle are ignored. This leads to a predicted quenching of emission and a reduction in quantum efficiency as shown in Figure \ref{Figure4}. On the other hand, the evaluated strengths of coupling with the metal particle shown in Figure \ref{Figure3}b indicate the breakdown of this assumption. The quenching is constant when the number of emitters increase in a constant area as $\rho$ = $\rho_c$ is constant. When $N$ increases and the system expands with a constant area density of emitters, this quenching due to the metal particle decreases as does the fraction of paths of emitted photons through the dissipative metal particle, with $\rho \longrightarrow 0$. The resulting quantum efficiency increases to approach $Q_0$ reflecting the density of states of a regular lattice of emitters. When some of the emitters in the model are replaced by metal particles (to reach a ratio of 1:6 for number of metal particles and emitters), quenching in this Markovian approximation is regained in the entire lattice with many coupled emitters and metal particles.

In case II, the Rabi oscillations among emitters as well as with the metal particle are included, and this leads to enhancements in the quantum efficiency (Figure \ref{Figure4}). The enhancement is constant as the number of emitters increase in a constant area, as both $\rho_c$ and $\rho_{nc}$ remain roughly constant. This shows the independence of LDOS to the number of such point emitters. To establish the correspondence of the LDOS to the density of optical states of a large lattice, we increase $N$ as the system is expanded with a constant area density of emitters. Here, $\rho_{nc} \longrightarrow 0$  due to weaker couplings, as does $\rho_c$. This results in a reduction of Rabi oscillations through the metal particle, and the efficiency approaches $Q_0$ of the lattice of emitters. The lost enhancement is regained when some of the emitters are replaced by metal nanoparticles fixing a ratio of 1:6 for the metal nanoparticles and the number of lattice sites.

The evaluations for the two geometries, and the two cases i.e. the Markovian and its one-loop correction, confirm that the correction presents physically meaningful predictions when the number of interacting emitters are many, and also with multiple metal particles. From the above two geometries studied, we can infer that the coherence of the paths among many strongly coupled emitters and metal nanoparticles is sustained independent of the number of emitters and metal particles. It should be noted that when the coupling of the emitter becomes stronger at small relative separations with the metal particle, this one-loop perturbation diverges and a full non-Markovian model of interaction becomes indispensable \cite{jain2021nonMarkov}. The above is also evident from the exponent of the proposed one-loop correction which is only first order in the coupling strength $g = \frac{\Omega}{\Gamma}$. Considering the lower dissipative loss of these very small metal particles they are expected to be much more effective in enhancement of spontaneous emission, compared to the highly scattering larger metal particles required in the weak-coupling regime. Further, the emerging dynamics in spontaneous emission from such materials can be exploited for applications other than light generation.

\renewcommand{\thefigure}{A\arabic{figure}}
\setcounter{figure}{0}



\section*{Appendix: Coupling strengths and size of gold nanoparticles} \label{B: Coupling strengths}
A larger factor $e^{-\Gamma/\Omega}$, determines the degree of divergence of observations from the predictions of the Markovian (weak-coupling) approximation of the emitter and the metal nanoparticle. Variation of this exponent in a logarithmic scale are plotted below in Figures \ref{FigureS12} and \ref{FigureS13} for a fixed small distance of 3 nm and for a ‘relative’ distance fixed as the radius ‘R’ of metal particle. All the above cases represent gold nanoparticles with a surrounding medium of refractive index 1.5, and at a free-space wavelength of 560 nm.

\begin{figure}[h!]
		\centering
		\includegraphics[width=1.0\linewidth]{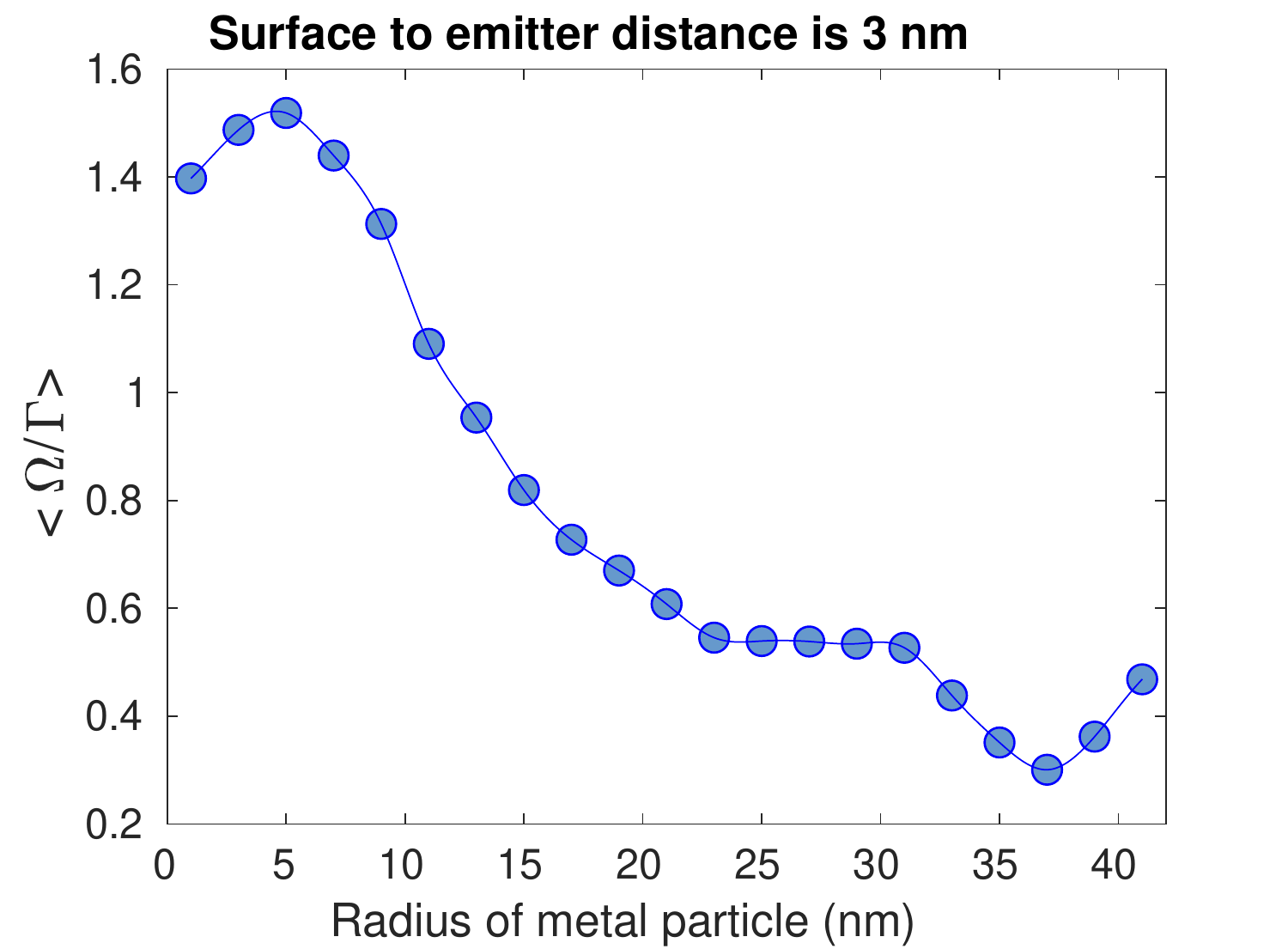}
		\caption{Polarization averaged coupling strengths of a single emitter and a gold nanoparticle as determined by the Markovian model, which can be used for one-loop corrections.}
		\label{FigureS12}
\end{figure}

\begin{figure}[h!]
		\centering
		\includegraphics[width=1.0\linewidth]{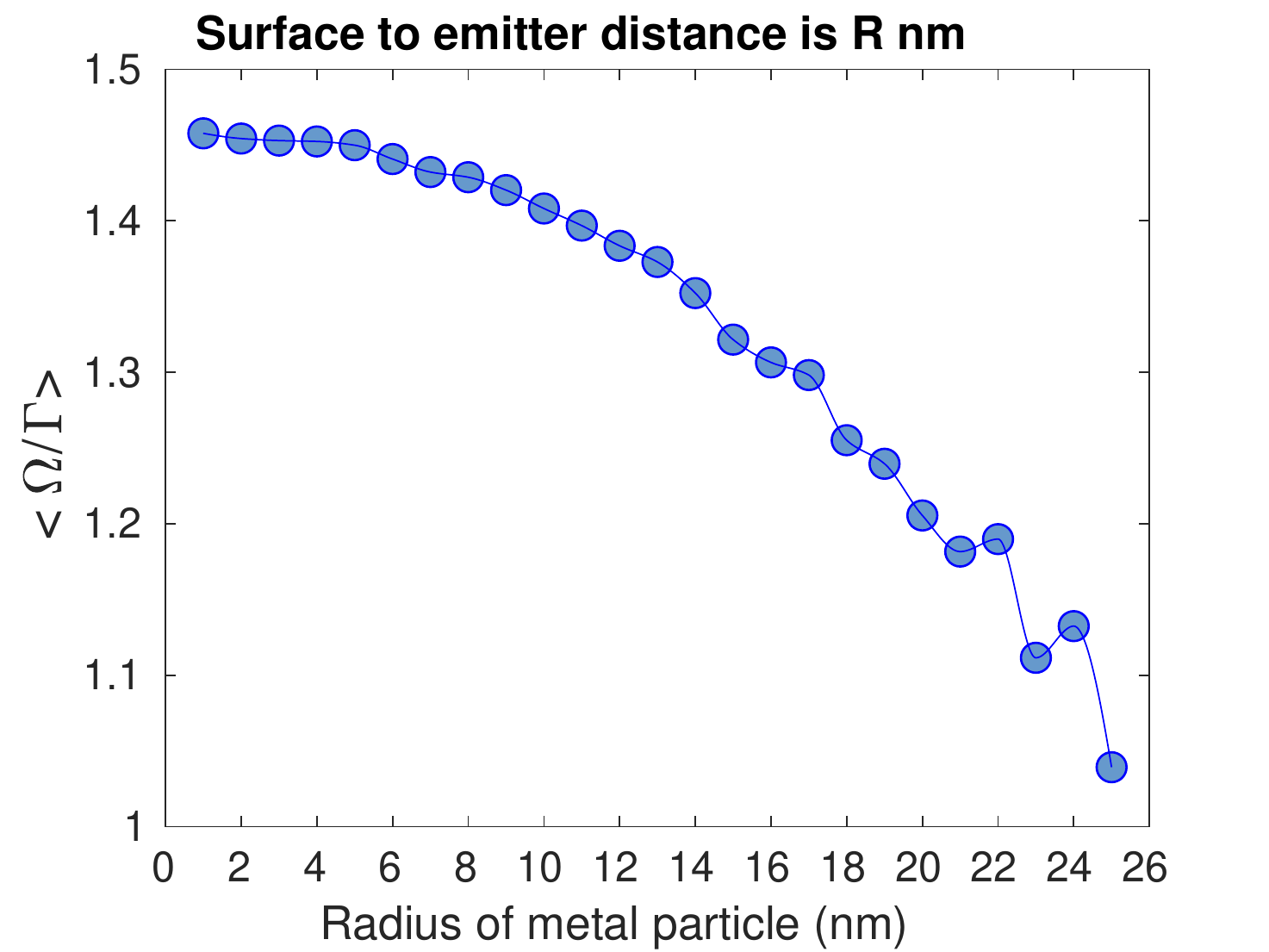}
		\caption{Polarization averaged coupling strengths of a single emitter and a gold nanoparticle as determined by the Markovian model, which can be used for one-loop corrections.}
		\label{FigureS13}
\end{figure}

\bibliography{references}
\end{document}